\documentclass[11pt,twoside]{article}

\usepackage{asp2006}
\usepackage{epsf}
\usepackage{lscape}
\usepackage{graphicx}

\begin{document}
\title{Merger History of Galaxies and Disk+Bulge Formation}   %%% Fill in title
\author{Sadegh Khochfar}   %%% Fill in author names
\affil{Max-Planck Institute for Extraterrestrial Physics, Giessenbachstrasse
D-85748 Garching, Germany}    %%% Fill in author affiliations

\begin{abstract} %%% Abstract to run on from here.
We discuss the transitions of galaxy morphologies within the CDM paradigm under the assumption of bulge formation in mergers and disk growth via cooling of gas and subsequent star formation. Based on the relative importance of these two competing processes it is possible to make predictions on the expected morphological mix of galaxies. In particular we here discuss the generation of massive disk galaxies with low bulge-to-total mass ratios. Our results indicate that it is difficult to generate enough massive disk galaxies with B/T $< 0.2$ via major mergers and subsequent disk re-growth, if during the major merger progenitor disks get disrupted completely. On average low B/T galaxies must have had there last major merger at $z \ge 2$. The main limiting factor is the ability to re-grow massive disks at late times after the last major merger of a galaxy. Taking into account the contribution from minor mergers  ( $4 \ge M_1/M_2$, $M_1 \ge M_2$) to the formation of bulges, we recover the right fraction of massive low B/T disk galaxies, indicating that minor mergers play an important role in the formation of massive low B/T disk galaxies.

\end{abstract}

%%% MAIN BODY OF TEXT GOES HERE. CONSULT "INSTRUCTIONS FOR AUTHORS USING
%%% LATEX2E MARKUP", SECTIONS 2.3-2.6 FOR HELP WITH EQUATIONS, FIGURES,
%%% AND TABLES.

%\section{}   %%% Top level section head (remove "%" symbol)
%\subsection{}   %%% Second level section head (remove "%" symbol)
%\subsubsection{}   %%% Lowest level section head (remove "%" symbol)
%\section*{}    %%% Unnumbered top level section head (remove "%" symbol)
%\subsection*{}   %%% Unnumbered second level section head (remove "%" symbol)
\section{Introduction}\label{intro}

The close resemblance of elliptical galaxies and {\it classical} bulges has 
lead to the widely accepted assumption that they have the same origin. 
Profiles of elliptical galaxies and bulges are nicely fit by  
Sersic-laws. The fact that super-massive black holes in bulges 
also follow the fundamental $M_{\bullet}$-$\sigma$-relation \citep{2001ApJ...550...65S} provides further evidence for a common formation scenario of elliptical  galaxies and classical bulges.

Early work by \cite{tt72} showed that elliptical galaxies can be the result of
a major merger between two spiral galaxies. 
Subsequent numerical simulations showed that indeed various properties of
elliptical galaxies and classical bulges can be recovered from simulations 
that use cosmological self-consistent initial orbital parameters \citep{kb06} 
for merging systems \cite[see e.g.][]{ba92,nb03,2006MNRAS.372..839N,2007MNRAS.376..997J}. 
As a consequence it should be possible to generalize results for 
the formation of elliptical galaxies to the formation of classical bulges 
and to speak more general of the formation of spheroids 
\citep{2006MNRAS.370..902K}.
 E.g. it has been predicted that massive spheroids
form in dry major mergers of elliptical galaxies, and that intermediate 
mass spheroids form as a result of a major merger between an elliptical and a 
spiral galaxy \citep{kb03,nk06}.  \cite{2006MNRAS.370..902K} find 
that this is indeed the case for ellipticals as well as bulges.
 
Bulges are embedded in large stellar disks in contrast to elliptical galaxies
which poses the question if they really can have the same origin. 
The $\Lambda$CDM paradigm offers a natural way for the transition from 
elliptical galaxies to bulges of early-type spirals via the accretion of a 
new disk in the aftermath of a major merger 
\citep{1999MNRAS.303..188K,sh05}. As \cite{kb01} show the predicted 
merger rate of galaxies in 
the $\Lambda$CDM paradigm is in fair agreement with the observed one which 
allows to test robustly the transition in Hubble types due to the growth of a 
new stellar disk. Hence the properties of bulges like e.g. the isophotal 
shape \citep{kb05} will initially be set by 
the properties of the progenitor elliptical galaxy.

\section{Model}
We use semi-analytical modeling of galaxy formation to predict the star burst 
and quiescent components of elliptical galaxies. The dark matter history is 
calculated using the merger tree proposed by \cite{som99} with a mass 
resolution of $2 \times 10^9 M_{\odot}$. The baryonic 
physics within these dark matter halos is calculated following recipes 
presented in \cite{kb05} and \cite{2006MNRAS.370..902K}. 
In our simulation, we assume that elliptical galaxies 
form whenever a major merger ($M_1 /M_2 \leq 4$ with $M_1 \geq M_2$) takes 
place. We assume that during this process all the cold gas which was in the
 progenitor disks will be consumed in  a central starburst, adding to the 
spheroid mass, and that all stars in the progenitor disks will be 
scattered into the spheroid too. Furthermore we allow the stars of satellite
 galaxies in minor mergers to also contribute to the spheroid.
 During the evolution of a galaxy, we keep track of the origins of all stars 
brought into the spheroid and attribute them to two categories, starburst and 
quiescent, where the first incorporates stars formed during a starburst 
in a major merger and the latter includes stars previously formed in a disk 
and added to the spheroid during a major merger. Each star will 
carry along its 
label and not change it, which means that if a star was made 
in a merger of two progenitor galaxies and the remnant of that merger 
participated in another merger, the star will  still contribute to the 
merger component of the final remnant.  For more modeling details, we refer 
the reader to \cite{2006MNRAS.370..902K} and references therein. Please note 
that our simulation does not include  
AGN-feedback (\cite{2006Natur.442..888S}) 
 or environmental effects \citep{2008ApJ...680...54K}  that have 
influence on the most massive galaxies.

\section{Results}
\begin{figure}
\center
\includegraphics[height=2.1in,width=2.6in,angle=0]{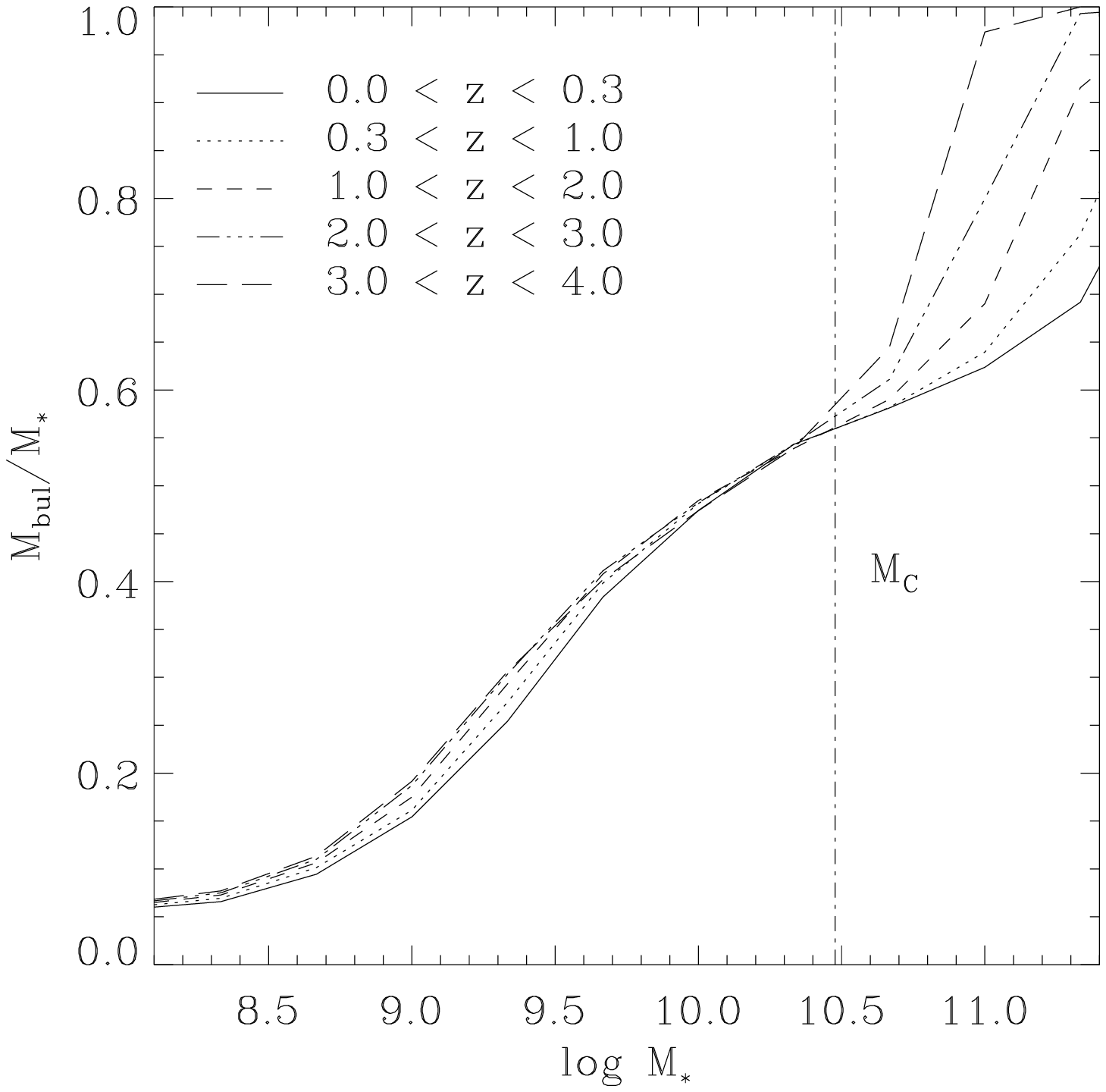}
\includegraphics[height=2.1in,width=2.6in,angle=0]{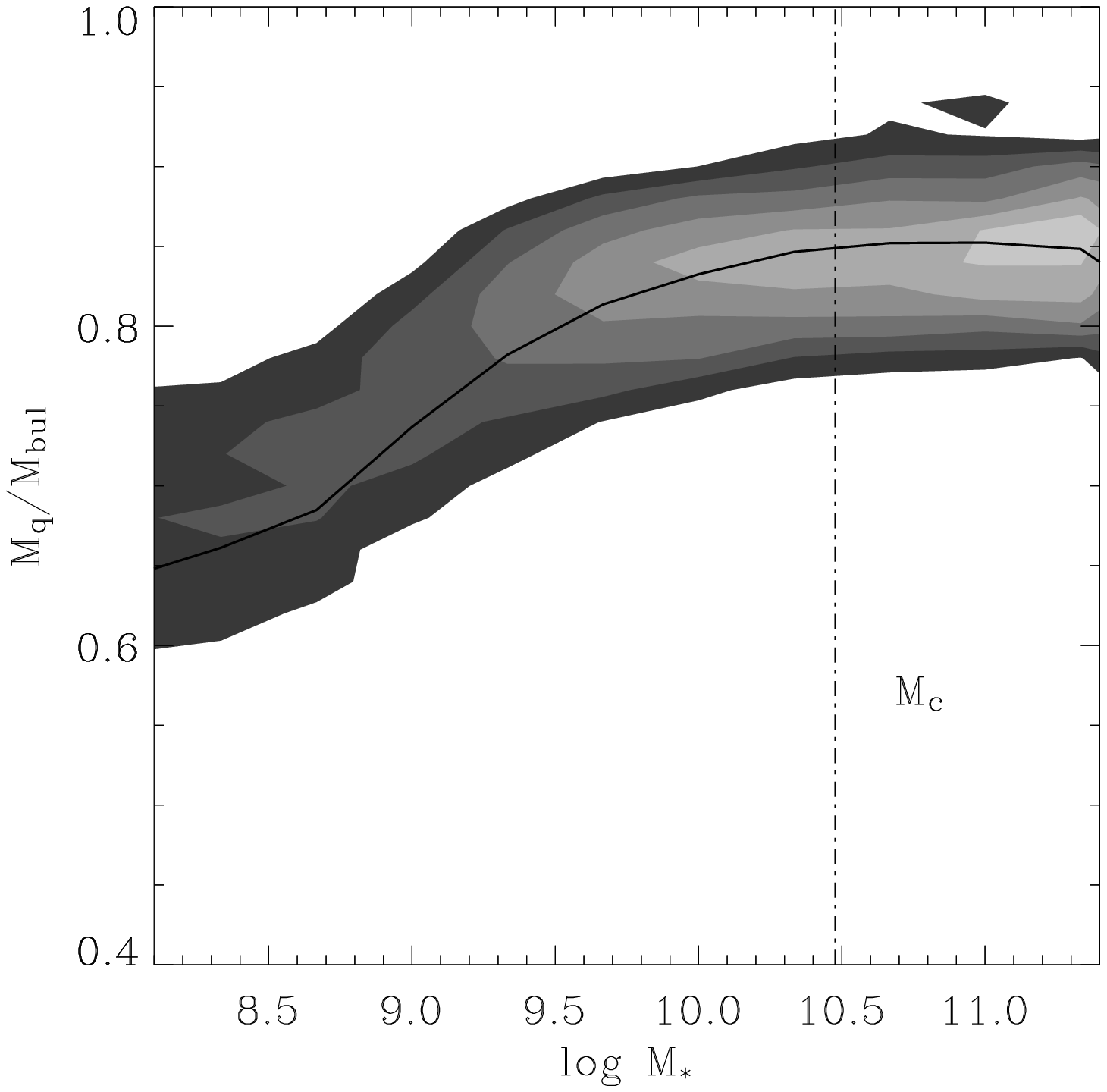}
  \caption{Left panel: Fraction of stars in galaxies of a given mass that 
reside in spheroids at various redshifts. Right panel: Quiescent fraction of 
stars in spheroids as a function of galaxy mass. The solid line shows the 
median of the distribution. The dot-dashed line indicates the critical mass 
scale $M_{\rm{c}}$} \label{f1}
\end{figure}
\subsection{Build-up of bulges}
Ongoing mergers constantly transfer disk stars to spheroids in the universe. 
If this process is more efficient than star formation in disks 
one is to expect an increase in the fraction of stars in spheroids over 
cosmic time. However, the merger rate is a strong decreasing function with 
redshift \citep{kb01} and at late times disk growth 
overtakes merging. In the left panel of Fig. 1 the fraction of 
stars in spheroids 
as a function of redshift and galaxy mass is shown. At early times the most 
massive galaxies, $M_* > M_{\rm{c}}\sim 3 \times 10^{10}$ M$_{\odot}$ 
are all elliptical galaxies and only at late times massive 
spiral galaxies appear. This is related to the gradual transformation of gas 
into stars in disks in contrast to the violent and fast transformation of 
gas into stars during major mergers and the ability of halos at high-z to accrete gas efficiently onto their central galaxy on approximately halo dynamical times \citep[][e.g.]{2005MNRAS.363....2K,2007ApJ...668L.115K,2008MNRAS.390.1326O,2008arXiv0812.1183K}. Many of the intermediate massive 
elliptical galaxies that formed at high redshift continue to grow disks to 
become bulges of present day spiral galaxies. 

The right panel of Fig. 1 shows the quiescent fraction of bulge stars 
as a function of galaxy mass. The quiescent fraction increases gradually until 
roughly $M_{\rm{c}}$ where it becomes constant at $\sim 0.85$. Most of the 
stars in bulges therefore originated from disks of progenitor galaxies or 
satellite galaxies. \cite{2006MNRAS.370..902K} report the number of 
minor satellite mergers exceeds that of  major mergers by an order of 
magnitude and is therefore one important driver for a high quiescent fraction 
in bulges. For massive bulges in addition mostly dry major mergers cause the 
quiescent fraction to  stay constant and not to change much \citep[see also][]{2008arXiv0809.1734K}, which explains 
the behavior at the high mass end.
 
Numerical simulations by \cite{sh05} show that dissipation accompanied by 
starbursts during major mergers leads to a population of stars that is more 
centrally concentrated than the scattered disk stars once they relaxed to 
a spheroid at the end of the merger. In our simulations we identified those 
centrally concentrated stars with the starburst component and the less 
concentrated previous disk stars with the quiescent component of bulges. 
\cite{2006ApJ...648L..21K} propose based on these two components a simple 
model in which the size of galaxies scales with the amount of dissipation 
during their formation and that is able to reproduce the size-evolution of 
early-type galaxies. In the left panel of Fig.~\ref{f2} we show the expected
size evolution of bulges as a function of their mass and formation time, i.e. 
we show the ratio of the present day effective radius of bulges, 
$r_{\rm{e,local}}$, to that of bulges at higher redshifts. 
Massive bulges that formed early are most likely to have had a significant 
amount of dissipation involved during their formation, because 
the gaseous disk only had enough time to transform a small portion of the 
gas into stars. In contrast the size-evolution for small bulges is not very 
strong, as there is only a small difference in the amount of dissipation.

The right panel of the same figure shows the quiescent fraction in bulges as a 
function of galaxy mass and environment. For galaxies more massive than 
$M_{\rm{c}}$ the quiescent fraction does not depend on the environment. 
Only for galaxies  below $M_{\rm{c}}$ we find an environmental dependence 
which reflects itself in a larger quiescent fraction for field galaxies. The 
reason for this is mainly buried in the larger amount of dissipation that is 
involved in the formation of bulges that end up in high density environments. 
These galaxies form in general earlier and therefore  
the amount of dissipation is larger during major mergers.

Observations of core phase-space densities in spiral galaxies 
reveal that they are several order of magnitudes lower than those of 
elliptical galaxies of the same mass 
\citep{1986ApJ...310..593C}. A possible solution to 
this problem is dissipation during starburst that can increase the phase space 
density in the remnant. If the centers of early-type 
spirals are dominated by bulges this suggest that bulges and ellipticals of 
the same mass must have had different amounts of dissipation during their 
formation. Indeed our simulations suggest that the quiescent fraction in 
bulges of spiral galaxies is higher than that of ellipticals of the same mass, 
which could explain the observations.

\begin{figure}
\center
\includegraphics[height=2.15in,width=2.7in,angle=0]{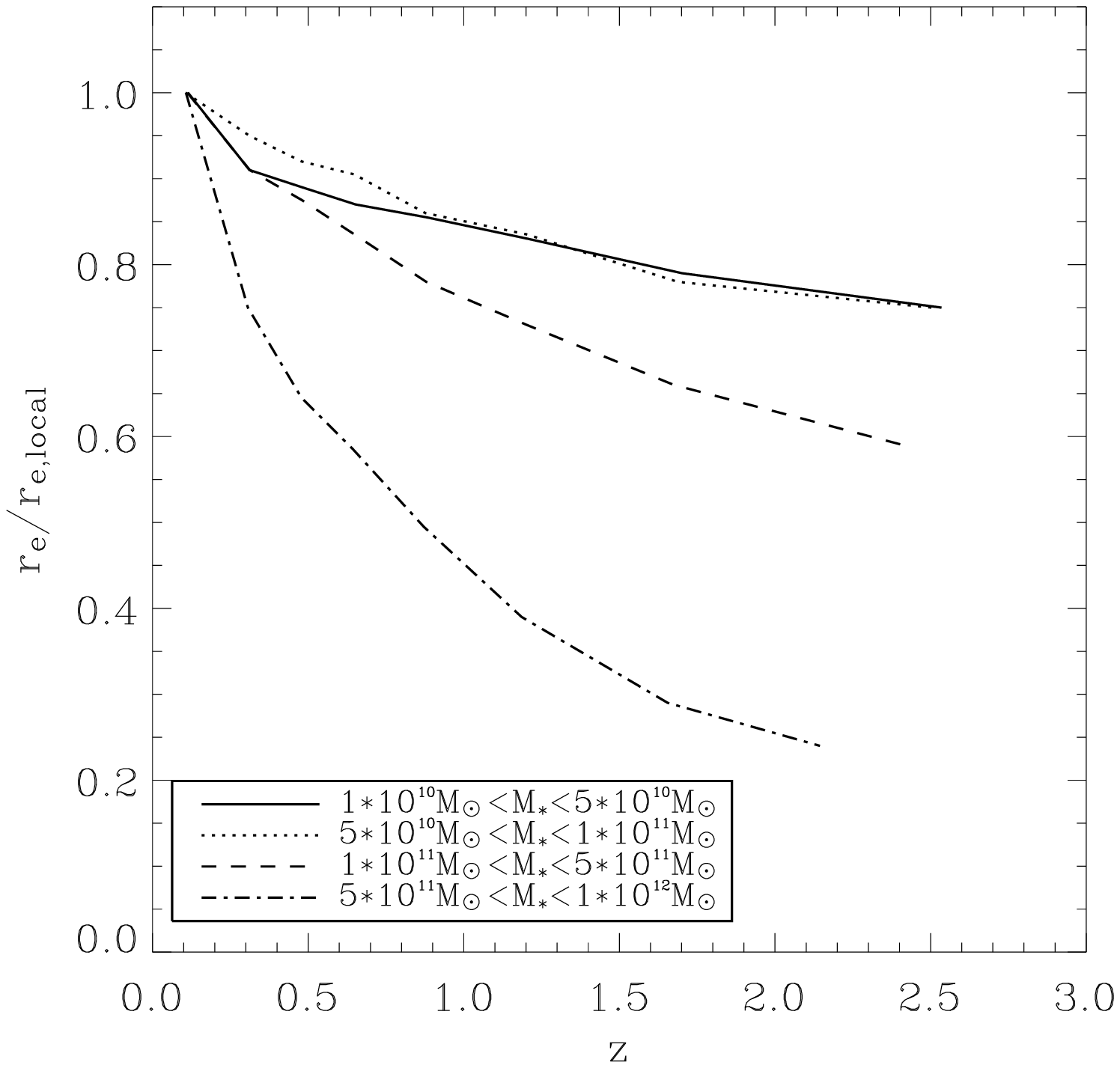}
\includegraphics[height=2in,width=2.5in,angle=0]{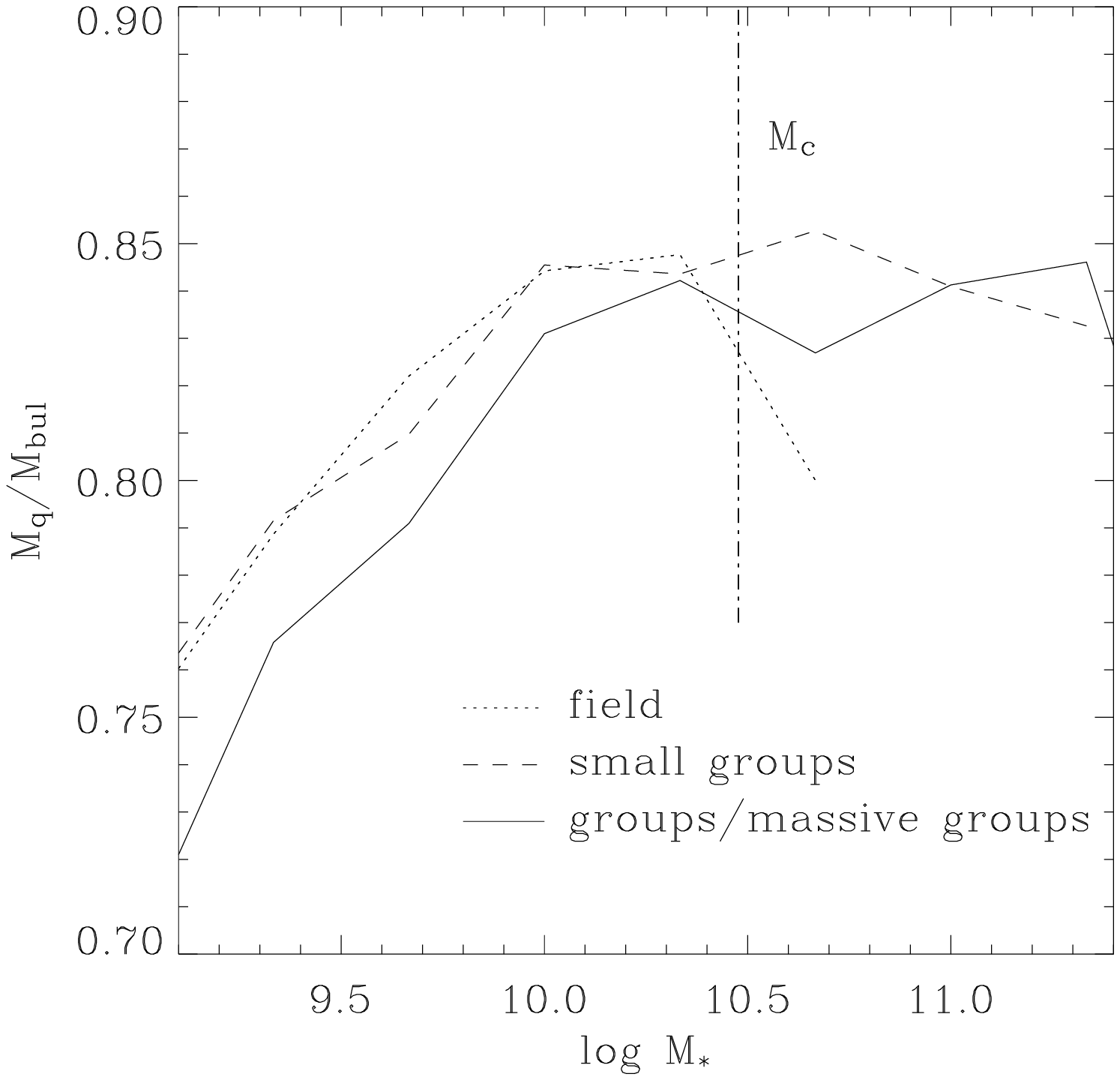}
  \caption{Left panel: size ratio between spheroids of the same mass at high 
    redshift and locally. Right panel: quiescent fraction of stars in 
    bulges as a function of galaxy mass and environment.  }\label{f2}
\end{figure}

\subsection{Bulge-to-total stellar mass ratios of disk galaxies}
Using above prescriptions it is  possible to investigate the distribution of bulge-to-total (B/T) stellar mass ratios within our SAM and compare them to recent observations of \cite{2008arXiv0807.0040W}. The left of Fig. 3 shows that our model  significantly underestimate the fraction of low B/T bulges for massive late-type galaxies with M$_* \ge 10^{10}$ M$_{\odot}$ if we only consider bulges that had at least one major merger during their history. The main reason for this result becomes clearer by looking at the B/T ratio of galaxies as a function of the time of their last major merger that led to the disruption of the disk (right part of Fig. 3). It appears that only  those galaxies that experienced a major merger before $z \ge 2$ significantly contribute to low B/T galaxies. These galaxies were the only ones that had enough time to re-grow a massive  disk after it was destroyed in the major merger. As mentioned earlier satellite mergers are by far more frequent than major mergers. In that respect they are able to contribute to the build up of bulges, especially low mass bulges. In our simple model we add the stars of the satellite to the bulge component of the host in minor mergers ($M_1/M_2 \ge 4$, $M_1 \ge M_2$). The small dotted line in the left panel  of Fig. 3  shows the prediction for bulges only build up via minor mergers. The majority of bulges made in this way live in low B/T $< 0.2$ bulges, while those made in major mergers have B/T $ \ge 0.2$. Adding the bulges that have experienced major mergers as well reproduces the overall distribution of observed bulges quite well. There are however, some interesting discrepancies. The fraction of large B/T galaxies is increasing too steep as a function of B/T compared to the observations. This can be traced back to possibly too efficient destruction of disk during major mergers in our model. Recent work by \cite{2009ApJ...691.1168H} showed that disks can actually survive major mergers if they are very gas-rich. This indeed could also explain the missing fraction of intermediate B/T galaxies that are missing in our current model. The main results however, that minor merger have a significant contribution to systems with low B/T also holds in a model with modified disk destruction \citep{2009arXiv0901.4111H}.
\begin{figure}
\center
\includegraphics[height=2.15in,width=2.7in,angle=0]{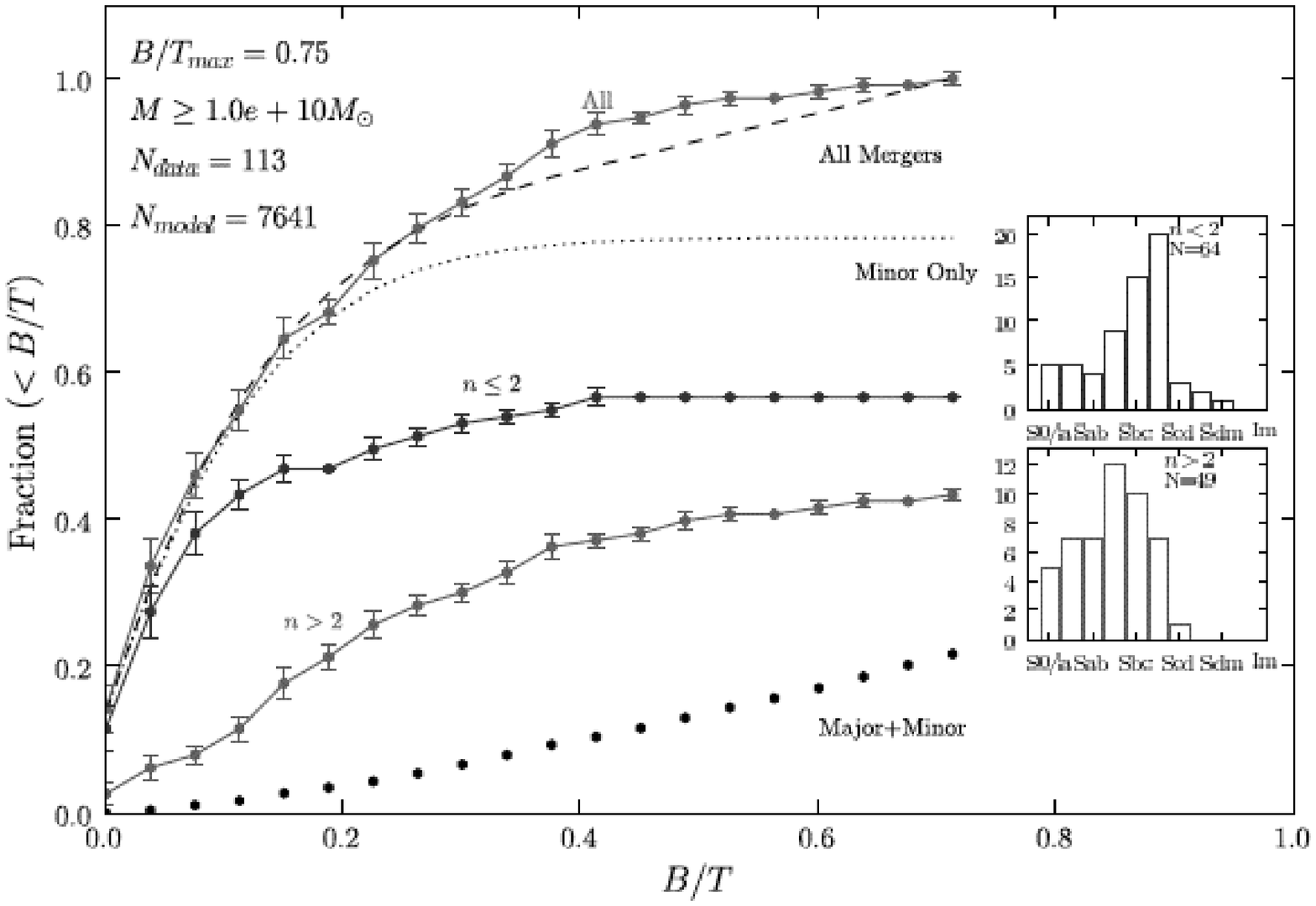}
\includegraphics[height=2in,width=2.5in,angle=0]{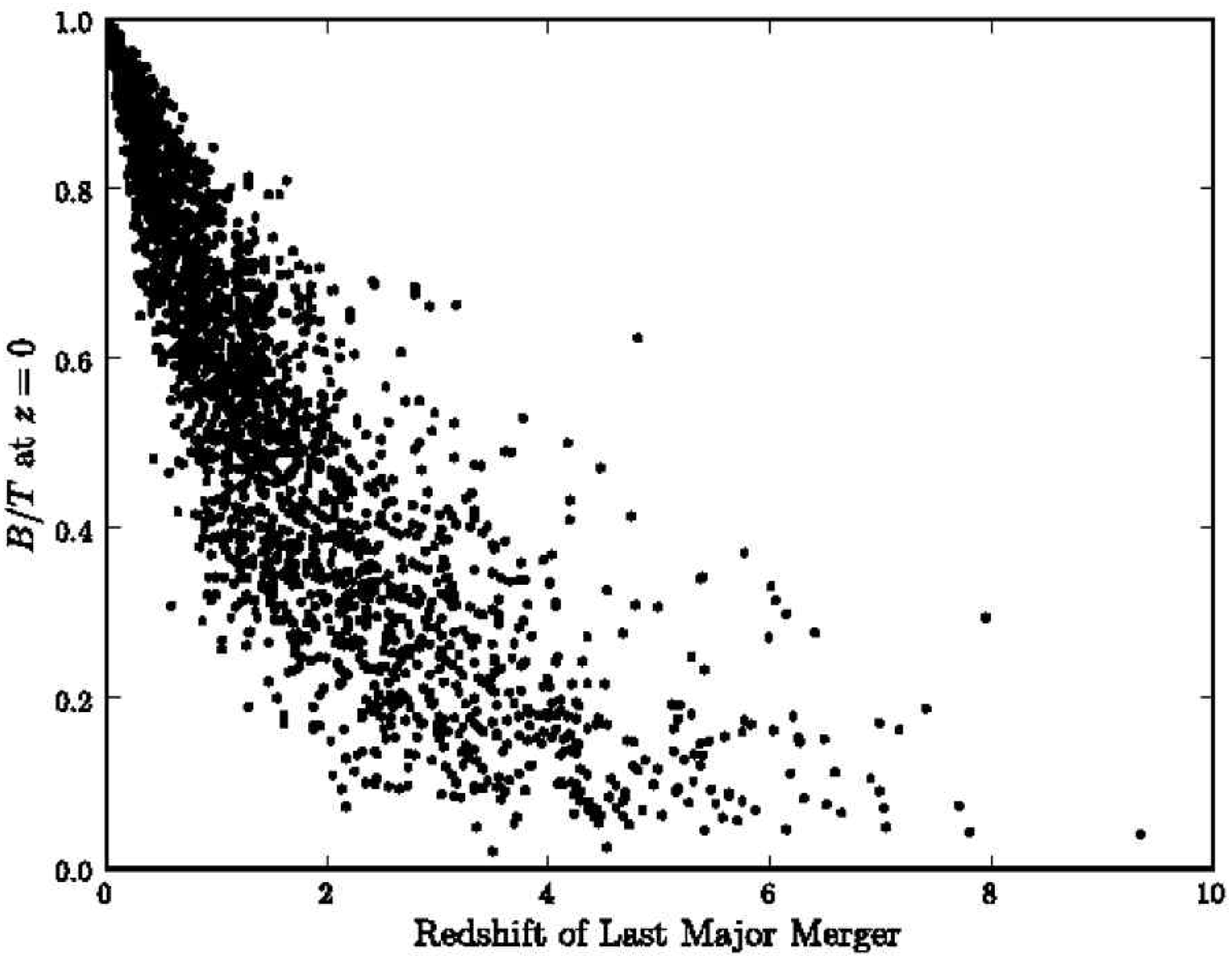}
  \caption{Left panel: Cumulative fraction of disk galaxies with bulge-to-total ratios less $B/T$. The small dotted line {\it minor only} refers to model bulges formed from only minor mergers and the big dotted line labeled {\it major+ minor} to bulges that had at least one major merger in their history. Solid black and grey line shows the observed disk galaxies with bulges having Sersic index $N \le 2$ and $n > 2$, respectively. Right panel: B/T of galaxies as a function of the redshift when they had their last major merger \citep[figures from][]{2008arXiv0807.0040W}  }\label{f2}
\end{figure}

\section{Conclusions}
In this paper we discussed how the merger history of galaxies affects their morphology in terms of disks and bulges. We showed that in a model where disks get disrupted in major mergers the majority of stars ending up in a bulge are actually made in progenitor disks. Only $15-40 \%$ of stars in present day spheroids, bulges and ellipticals, are made from starbursts in gas-rich mergers.
Thus we would argue that the main mode of star formation is in disks and not in starbursts triggered by galaxy interactions. Furthermore, we find that the fraction of stars made in starburst increases as a function of redshift due to dissipation playing a more important role at high z. Cooling times of gas are short resulting in large quantities of unprocessed gas in progenitor disks that can be transformed into stars during major mergers. Taking into account the amount of dissipation during merger and the amount of satellite mergers it is possible to recover the size-evolution of early-type galaxies with time. Massive early-type galaxies at low redshift have had many mergers with satellite galaxies that puffed them up to become larger in size over time. In contrast massive early-type galaxies at high redshift did form from a single very gas-rich merger leaving behind a compact remnant.  

When considering the growth of bulges from major mergers we find that the fraction of massive disk galaxies with B/T $< 0.2$ is far too low compared to the observations. In addition the fraction of galaxies with B/T $ > 0.2$ increases to steep as function of B/T. Latter is possibly due to a too simplified model for the effects on progenitor disks by major mergers. Our model assumes a total destruction which has been recently shown to be not necessarily the case. As a consequence we might underestimate intermediate B/T ratios and overestimate high B/T ratios. Independent of that however, we find that the bulges of the B/T $ < 0.2$ population of disk galaxies can be associated with a series of minor mergers forming them in our model.  Assuming that satellites contribute their stars to the bulge of the host galaxy we find a steep rise in the fraction of low B/T galaxies in agreement with recent observations. Combining both, bulges formed by minor mergers only and those by at least one major merger and a series of minor mergers, we recover the overall trend in the fraction of disk galaxies as a function of B/T. Though these are promising results future work will have to show what the role of minor mergers in the formation of bulges is, and how it reflects on other properties like e.g. the bar fraction in disk galaxies.

\acknowledgements %%% Text of acknowledgements runs on after this command.

I would like to thank Tim Weinzirl \& Shardha Jogee for many useful discussions.

%%% THE BIBLIOGRAPHY
%%%
%%% CONSULT SECTION 3 OF "INSTRUCTIONS FOR AUTHORS" FOR HOW TO USE NATBIB.
%%% AUTHORS ARE ENCOURAGED TO USE EITHER THE "THEBIBLIOGRAPY" ENVIRONMENT
%%% BY UNCOMMENTING (DELETING THE "%" SYMBOL) THE COMMANDS BELOW, OR BY
%%% USING THE BIBTEX ENVIRONMENT. TO FIND OUT WHICH IS APPLICABLE TO YOUR
%%% CONTRIBUTION, CONSULT THE VOLUME EDITORS FOR YOUR PROCEEDINGS.
%%%

\end{document}